\documentclass[sigconf]{acmart}

\usepackage{enumitem}

\usepackage{multirow}
\usepackage{makecell}
\usepackage{prettyref}
\newrefformat{fig}{Figure~\ref{#1}}
\AtBeginDocument{%
  }

\title{UniScale: Synergistic Entire Space Data and Model Scaling for Search Ranking}

\author{Liren Yu}
\authornote{These authors contributed equally to this work.}
\affiliation{%
  \institution{Taobao $\&$ Tmall Group of Alibaba}
  \city{Hangzhou}
  \country{China}
}
\email{yuliren.ylr@taobao.com}

\author{Caiyuan Li}
\affiliation{%
 \institution{Taobao $\&$ Tmall Group of Alibaba}
  \city{Hangzhou}
  \country{China}}
\email{licaiyuan.lcy@taobao.com}
\authornotemark[1]

\author{Feiyi Dong}
\affiliation{%
 \institution{Taobao $\&$ Tmall Group of Alibaba}
  \city{Hangzhou}
  \country{China}}
\email{dongfeiyi.dfy@taobao.com}
\authornotemark[1]

\author{Tao Zhang}
\authornote{Corresponding author.}
\affiliation{%
 \institution{Taobao $\&$ Tmall Group of Alibaba}
  \city{Beijing}
  \country{China}
}
\email{quen.zt@alibaba-inc.com}

\author{Zhixuan Zhang}
\affiliation{%
 \institution{Taobao $\&$ Tmall Group of Alibaba}
  \city{Hangzhou}
  \country{China}}
  \email{zhibing.zzx@taobao.com}

\author{Dan Ou}
\affiliation{%
 \institution{Taobao $\&$ Tmall Group of Alibaba}
  \city{Hangzhou}
  \country{China}}
  \email{oudan.od@taobao.com}

\author{Haihong Tang}
\affiliation{%
 \institution{Taobao $\&$ Tmall Group of Alibaba}
  \city{Hangzhou}
  \country{China}}
  \email{piaoxue@taobao.com}

  \author{Bo Zheng}
\affiliation{%
 \institution{Taobao $\&$ Tmall Group of Alibaba}
  \city{Beijing}
  \country{China}}
  \email{bozheng@alibaba-inc.com}

\begin{document}




\begin{abstract}
  Recent advances in Large Language Models (LLMs) have inspired a surge of scaling research in industrial search, advertising, and recommendation systems. However, existing approaches focus mainly on architectural improvements, overlooking the critical synergy between data and architecture design. We observe that scaling model parameters alone exhibits diminishing returns, and that the performance degradation caused by complex heterogeneous data distributions is often irrecoverable through model design alone. In this paper, we propose UniScale, a novel co-design framework that jointly optimizes data and architecture to unlock the full potential of model scaling. UniScale includes two core parts: (1) ES³ (Entire-Space Sample System), a high-quality data scaling system that expands the training signal beyond conventional sampling strategies through intra-domain expansion with hierarchical label attribution and cross-domain searchification; and (2) HHSFT (Heterogeneous Hierarchical Sample Fusion Transformer), a novel architecture that effectively models the complex heterogeneous distribution of scaled data via Heterogeneous Hierarchical Feature Interaction and Entire Space User Interest Fusion, thereby surpassing the performance ceiling of structure-only model tuning.
  Extensive experiments on large-scale industrial datasets demonstrate that UniScale achieves significant improvements and exhibits clear scaling trends. Online A/B tests on a real-world E-commerce search platform confirm that UniScale consistently outperforms strong production baselines, achieving 1.70\% and 2.04\% increases in user purchase and Gross Merchandise Value (GMV).
\end{abstract}

\ccsdesc[500]{Information systems~Retrieval models and ranking}
 

\keywords{Industrial Search System, Entire Space Modeling, Scaling Laws, Feature Interaction}

\maketitle

\section{Introduction}

Recommendation and search systems serve as indispensable channels for users to access information in modern internet ecosystems such as e-commerce, streaming media, and social networks~\cite{covington2016deep,zhou2018deep,naumov2019deep,lian2022persia,chang2023pepnet,pancha2022pinnerformer,xia2023transact,zhai2024actions,zhang2024wukong,lin2025can,liu2022monolith,yu2025hhft}. Due to the inherent complexity of user behaviors and the diversity of system inventory, precisely estimating user preferences remains a significant challenge in large-scale industrial systems. State-of-the-art approaches rely on deep learning-based ranking models~\cite{covington2016deep,cheng2016wide,guo2017deepfm,zhou2018deep,wang2021dcn}. Recently, inspired by the success of Large Language Models (LLMs)~\cite{vaswani2017attention,hoffmann2022training,kaplan2020scaling,achiam2023gpt}, the community has begun scaling ranking models through architectural upgrades and larger parameter budgets~\cite{zhang2024wukong,zhai2024actions,zhu2025rankmixer}. Nevertheless, pure parameter scaling exhibits diminishing marginal gains: simply stacking more parameters fails to yield further improvements. We argue that this bottleneck is fundamentally a \textit{data} bottleneck rather than a modeling one---under a fixed data volume, even highly expressive models cannot fully realize their potential. Two empirical observations in our study support this view: (i) naively expanding training data on a standard backbone causes a $-0.43\%$ AUC degradation, while a co-designed architecture converts the same data into a $+0.32\%$ AUC gain (Table~\ref{tab:sft}), so data composition is itself a first-order constraint on what model scaling can deliver; (ii) the AUC gap between search-only data and entire-space data widens monotonically with model scale, from $+0.12\%$ at 40M parameters to $+0.32\%$ at 300M parameters (Figure~\ref{fig:scaling_exp}), echoing compute-optimal observations in language-model scaling where data and parameters must grow jointly~\cite{hoffmann2022training}.

In this paper, we propose \textbf{UniScale}, a co-design framework that jointly optimizes data and architecture to unlock the full potential of model scaling. UniScale includes two core parts: (1) \textbf{ES³ (Entire-Space Sample System)}, a high-quality data scaling system that expands training signals beyond conventional sampling; and (2) \textbf{HHSFT (Heterogeneous Hierarchical Sample Fusion Transformer)}, an architecture that effectively models the complex heterogeneous distribution of scaled data, surpassing the ceiling of structure-only tuning. Although our deployment is on a search ranking system, the methodology is general to industrial ranking pipelines: intra-domain uniform sampling applies to any retrieval funnel, hierarchical label attribution applies to any multi-stage system with cross-channel interactions, and cross-domain searchification applies whenever multiple interaction domains coexist.

Existing industrial practices typically expand training samples via heuristic augmentation~\cite{xie2022contrastive} or negative sampling~\cite{huang2020embedding,yang2020mixed,mu2023hybrid}, yet the degradation from complex heterogeneous data is often irrecoverable through model design alone. In search systems, ranking is essentially a fine-grained selection among query-relevant candidates, which imposes stringent quality requirements on any additional data introduced during scaling.

To this end, ES³ scales the data by constructing a high-quality, entire-space training set through two mechanisms: \textbf{(1) Intra-domain Sample and Label Expansion}, which incorporates unexposed items from the same request contexts as supplementary data to mitigate selection bias, and introduces a Hierarchical Label Attribution mechanism that propagates user feedback signals from the entire interaction space to both exposed and unexposed samples (delayed cross-domain conversions are also treated as complementary labels); \textbf{(2) Cross-domain Sample Searchification}, which transforms non-search user-item interactions into search-aligned training samples through a \textit{Negative Sample Generation Module} and a \textit{Feature Alignment Module}, preserving semantic coherence between the pseudo-search request and synthetic query, with all transformed samples flowing through the search feature logging infrastructure for full schema alignment.

The resulting dataset exhibits significantly richer and more complex distributions than the conventional setting, demanding a more powerful architecture. We accordingly propose HHSFT with two modules: \textbf{(1) Heterogeneous Hierarchical Feature Interaction}, which partitions features into semantic blocks, tokenizes them, applies token-specific projections and FFNs, then aggregates via composite projection to model high-level interactions; and \textbf{(2) Entire Space User Interest Fusion}, which uses a Domain-Routed Expert Fusion layer to decouple cross-domain shared and domain-specific knowledge, plus a Domain-Aware Personalized Gated Attention to selectively inject cross-domain user interests into search ranking.

Finally, to deploy UniScale under strict latency and computational constraints, we implement optimizations including feature pre-hashing, attention kernel fusion, RDMA, and fp16 quantization. Importantly, ES³ is purely a training-time data construction system and does not introduce additional online serving computation; its only overhead is extra storage for the expanded samples, which is small compared to online compute resources, yielding a high ROI in production. The contributions of this paper are:
\begin{itemize}[leftmargin=*, noitemsep]
\item We identify fundamental limitations of conventional sampling strategies in industrial ranking and propose UniScale, a co-design framework that jointly optimizes data and architecture to break through the performance bottleneck caused by limited information and biased model learning.
\item We propose ES³, an Entire-Space Sample System enhancing data from both intra-domain and cross-domain user feedback with mitigated negative transfer, and HHSFT, including heterogeneous feature tokenization, hierarchical feature interaction, and entire-space user interest fusion to model complex distributions.
\item We conduct extensive offline and online experiments on billion-scaled industrial data, verifying the importance of co-design of data and architecture. UniScale has been deployed in Taobao Search Ranking System, achieving 1.70\% and 2.04\% increases in user purchase and Gross Merchandise Value (GMV).
\end{itemize}

\section{Related Work}

\subsection{Scaling up Industrial Ranking Models}
The evolution of recommendation systems is increasingly governed by scaling laws—model performance improves predictably with parameters, data volume, and compute. Recent industrial frameworks like OneRec~\cite{deng2025onerec} and OneSearch~\cite{chen2025onesearch} demonstrate large-scale pre-training power. HSTU~\cite{zhai2024actions} improves sequence-length scalability via a spatially efficient transformer. Wukong~\cite{zhang2024wukong} explores scaling laws for explicit high-order feature interactions. RankMixer~\cite{zhu2025rankmixer} enables scalable high-order interactions through token-mixing. OneTrans~\cite{zhang2025onetrans} provides unified Transformer backbones for sequential and non-sequential inputs. However, the returns from pure parameter scaling often exhibit diminishing marginal gains. We attribute this bottleneck to the limited information capacity of the training data: with data volume held constant, the full potential of highly expressive models remains unrealized. We therefore propose a co-design framework that jointly optimizes data and architecture.

\subsection{Entire Space Sample Modeling}
Data sparsity and sample selection bias necessitate holistic modeling beyond observed exposures. ESMM~\cite{ma2018entire} pioneers implicit entire-space modeling, later extended to multi-domain settings (STAR~\cite{sheng2021one}, PEPNet~\cite{chang2023pepnet}). However, direct fusion of heterogeneous domains often induces negative transfer.
While MoE architectures (MMoE~\cite{ma2018modeling}, PLE~\cite{tang2020progressive}) mitigate task conflicts via soft gating, they remain vulnerable to the seesaw phenomenon. Sparse MoEs with hard routing (Expert Choice~\cite{zhou2022mixture}, Switch Transformer~\cite{fedus2022switch}) scale well in LLMs but lack user-level interest fusion. Differing from these methods, we first scale data by constructing a high-quality, entire-space training set and then build a hierarchical multi-expert network to fully fuse entire-space user interest.

\section{Method}

\subsection{Problem Formulation}

Given a user $ u \in \mathcal{U} $, an item $v\in \mathcal{I} $, contextual information $ c \in \mathcal{C} $ (e.g., query, time, device) and a domain identifier  $d\in \mathcal{D}$ (e.g., search, recommendation, ad campaign), and the user feedback label $y \in \{0,1\}$, the user feedback prediction task aims to estimate the probability $ p(y=1|u, v, c, d) $ that user $ u $ will interact with item $ v $ under context $ c $ in domain $d$. Formally, we define the input feature vector as $ \mathbf{x} = [{u}; {v}; {c}; {d}]\in \mathbb{R}^{d_X}$. Given the training set $\mathcal{S}=\cup_{d\in \mathcal{D}}\mathcal{S}_d=\cup_{d\in \mathcal{D}} \{(\mathbf{x},y)\}_d$, the goal is to learn a model to predict the probability that
$
\hat{y}  \approx p(y=1|\mathbf{x})
$.

\begin{figure}[!ht]
  \centering
  \vspace{-6pt}
  \includegraphics[width=1.0\linewidth]{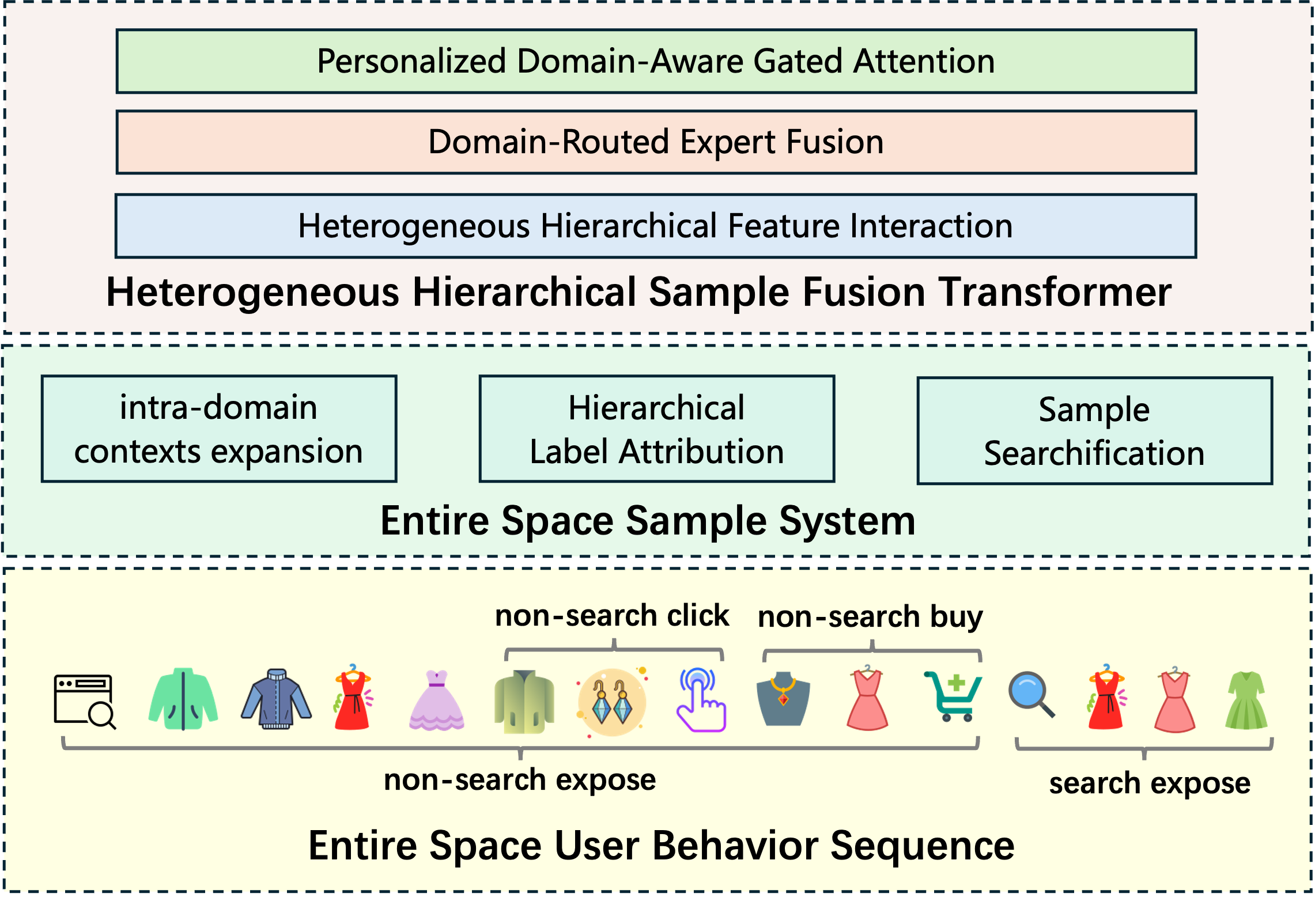}
  
\vspace{-10pt}
  \caption{Overall Architecture of Proposed UniScale, which combines Entire Space Sample System (ES³) for high quality data scaling and Heterogeneous Hierarchical Sample Fusion Transformer (HHSFT) for fully incorporating user interests expressed by entire space user behavior.}
  \label{fig:overview} 
\vspace{-12pt}
\end{figure}

\subsection{Overall Architecture}
As illustrated in Figure~\ref{fig:overview}, UniScale consists of two parts: (a) the Entire Space Sample System (ES³) for high-quality data scaling, comprising intra-domain sample expansion with hierarchical label attribution and cross-domain sample searchification; and (b) the Heterogeneous Hierarchical Sample Fusion Transformer (HHSFT) for incorporating entire-space user interests, comprising heterogeneous hierarchical feature interaction and entire-space user interest fusion.

\subsection{Entire-Space Sample System (ES³)}

\begin{figure}[!ht]
  \centering
  \vspace{-6pt}
  \includegraphics[width=1.0\linewidth]{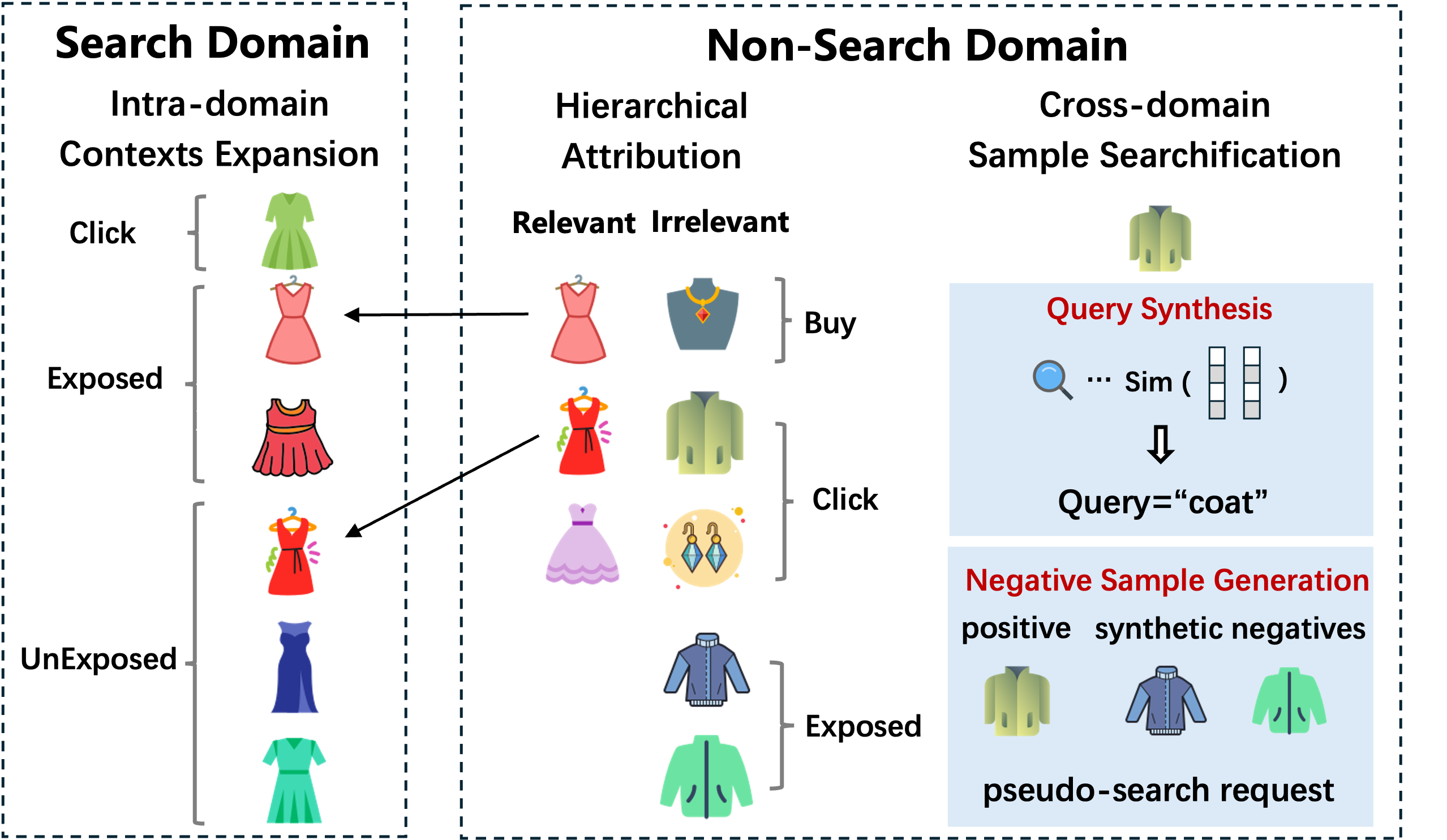}

\vspace{-10pt}
  \caption{Overview of the Entire-Space Sample System (ES\textsuperscript{3}). ES\textsuperscript{3} constructs a unified, bias-mitigated training set by synergistically integrating samples across the entire user behavior space. The framework comprises two core modules: (i) Intra-domain Sample and Label Expansion; (ii) Cross-domain Sample Searchification. }
  \label{fig:es3}
\vspace{-12pt}
\end{figure}

Ranking models in search systems are conventionally trained on exposed samples with in-domain labels, which induces three structural bottlenecks that become more pronounced as model capacity grows: \textit{selection bias} (unexposed candidates lack supervision and create a self-reinforcing exposure loop), \textit{label sparsity} (cross-domain feedback is discarded by single-domain attribution, leaving tail items poorly supervised), and \textit{cross-domain blind spots} (non-search interactions are entirely excluded due to schema heterogeneity).
The proposed \textbf{Entire-Space Sample System (ES\textsuperscript{3})} (Figure~\ref{fig:es3}) addresses each bottleneck by a dedicated mechanism: \textit{(i) intra-domain unexposed sample expansion} (Section~\ref{subsec:unpv}) targets selection bias by uniformly sampling from the full candidate list; \textit{(ii) Hierarchical Label Attribution} (Section~\ref{subsec:unpv}) targets label sparsity by propagating cross-domain feedback to both exposed and unexposed search samples ($2\times$ positive labels, $3\times$ samples, see Table~\ref{tab:es3}); \textit{(iii) Cross-domain Sample Searchification} (Section~\ref{subsec:feisou}) targets blind spots by transforming non-search interactions into schema-aligned pseudo-search samples. Mechanisms (i)--(ii) form the intra-domain stream and (iii) forms the cross-domain stream, jointly producing an entire-space training set that is both denser and broader than conventional sampling.

\subsubsection{Intra-Domain Sample and Label Expansion}\label{subsec:unpv}

Traditional ranking models trained exclusively on exposed samples suffer from two intertwined limitations. First, a critical \textit{training-inference mismatch}: exposed-but-unclicked items serve as hard negative examples within the training space, while the absence of easy negatives in training leads to prediction inaccuracy on the full candidate set. Second, \textit{label sparsity and exposure bias}: click labels are bound to real-time exposure logs, while conversion labels adopt last-click attribution, which assigns conversion credit solely to the user's final search click on the item. This paradigm fragments cross-scenario signals, yielding sparse and biased interest representations.

\begin{itemize}[label={},leftmargin=0pt, noitemsep]
\item  \textbf{Intra-Domain Unexposed Samples Expansion.} To bridge the training-inference gap, we incorporate unexposed samples via random sampling from the full candidate list of each search request. This expansion aligns the training distribution with the inference space, significantly enhancing the model's discriminative capability across the entire candidate set and mitigating selection bias.
\end{itemize}

However, directly integrating unexposed samples as pure negatives is problematic. These items inherently possess negative in-domain labels, which would exacerbate popularity bias. Meanwhile, even for originally exposed samples, the in-domain labels only capture fragmented search-session feedback. A user may be exposed to an item in search without immediate interaction, yet later convert via off-search pathways (e.g., recommendation feed or item detail page). Such valuable positive signals remain unutilized under conventional attribution. To address both issues simultaneously, we introduce a hierarchical label attribution mechanism that enriches supervision for the entire intra-domain sample space.

\begin{itemize}[label={},leftmargin=0pt, noitemsep]
\item  \textbf{Hierarchical Label Attribution.}\label{subsec:guiyin} We propose a hierarchical attribution paradigm to systematically recover cross-scenario behavioral signals with unified priority logic:
\end{itemize}

\begin{itemize}[leftmargin=*, noitemsep, topsep=0pt]
    \item \textit{Cross-domain clicks}: Attribute in descending priority to (1) search-exposed samples, (2) unexposed samples;
    \item \textit{Cross-domain conversions}: Attribute in descending priority to (1) search click samples, (2) exposed-but-unclicked samples, (3) unexposed samples.
\end{itemize}

This paradigm bidirectionally enriches the intra-domain sample space, and---more importantly---\textit{corrects} an existing systematic bias rather than introducing a new one. The conventional ``unexposed = uninterested'' label is itself an artifact of past ranking decisions: items remain unexposed because the system chose not to surface them, not because users genuinely dislike them. Models trained day after day on such labels assign progressively lower scores to unexposed and tail items, forming a self-reinforcing feedback loop that further suppresses them. Cross-domain conversion signals---in particular high-confidence ones such as same-day purchases on item detail pages---are generated outside the search exposure mechanism and therefore constitute an independent measurement of user interest. Propagating them yields two complementary effects: (a) for unexposed samples, an implicit-negative artifact is replaced with an authentic positive, breaking the loop and providing genuine supervision for tail items; (b) for originally exposed samples, delayed cross-domain conversions act as complementary positive labels ($2\times$ positive labels), densifying tail supervision. We acknowledge that propagated samples introduce a distributional discrepancy between source and target domains; this is precisely what HHSFT's DREF and DAPGA modules (Section~\ref{sec:moe_layer}) are designed to absorb structurally, rather than relying on label-level noise tolerance. The entire attribution pipeline runs offline on a daily cadence and adds no online serving latency.

\subsubsection{Cross-domain Sample Searchification}\label{subsec:feisou}

A substantial volume of user interactions in Taobao exhibit no explicit search association yet carry valuable interest signals. Direct adoption confronts two challenges: (1) sample-distribution divergence from the search scenario, and (2) the absence of search query fields and feature-schema heterogeneity. We address them with a Sample Searchification Engine consisting of two components:

\noindent\textbf{Negative Sample Generation Module.} Non-search requests contain semantically heterogeneous items (e.g., apparel and electronics in the same feed); naively treating exposed-but-unclicked items as negatives would inject query-irrelevant noise. We instead apply similarity-aware sampling: for each non-search click sample, we select unclicked items semantically similar to the clicked item from the same-request exposure sequence as synthetic negatives, preserving semantic consistency with the synthetic query.

\noindent\textbf{Feature Alignment Module.} To fill in the missing query field, we apply a hierarchical strategy: (i) historical query reuse from the user's prior searches for the item; (ii) item--query co-occurrence statistics with significance thresholds; (iii) ANN-based semantic retrieval over item title embeddings and the query corpus. Non-search click samples and their synthetic negatives then flow through the search feature logging infrastructure, yielding instances with feature schemas fully aligned to original search samples.

Quality of the synthesized samples is enforced by construction: historical query reuse only draws from real user-clicked queries; co-occurrence selection requires statistical significance thresholds; ANN retrieval enforces a minimum cosine similarity. Combined with the hierarchical attribution in Section~\ref{subsec:guiyin}, ES\textsuperscript{3} comprehensively covers cross-scenario user behaviors.

\noindent\textbf{Operational Safeguards.} For production reliability, ES\textsuperscript{3} runs a multi-layered safeguard: (1) \textit{data quality}---feature coverage, label distribution, and sample freshness are tracked on synthesized samples; (2) \textit{model monitoring}---AUC, prediction-score means, and outlier counts continuously surface degradation; (3) \textit{staged rollout}---pipeline changes go through controlled A/B tests before full deployment. A dedicated serving-side calibration module keeps predictions stable under distribution shifts (orthogonal to our contribution, hence omitted here).

To handle the enlarged sample space, we reformulate data list-wise---each row holds one request and its candidates---reducing redundant user/query storage ($\sim$50\% storage and IO) and computing user/query features once per request, improving training throughput.

\begin{figure*}[!ht]
  \centering
  \includegraphics[width=1.0\linewidth]{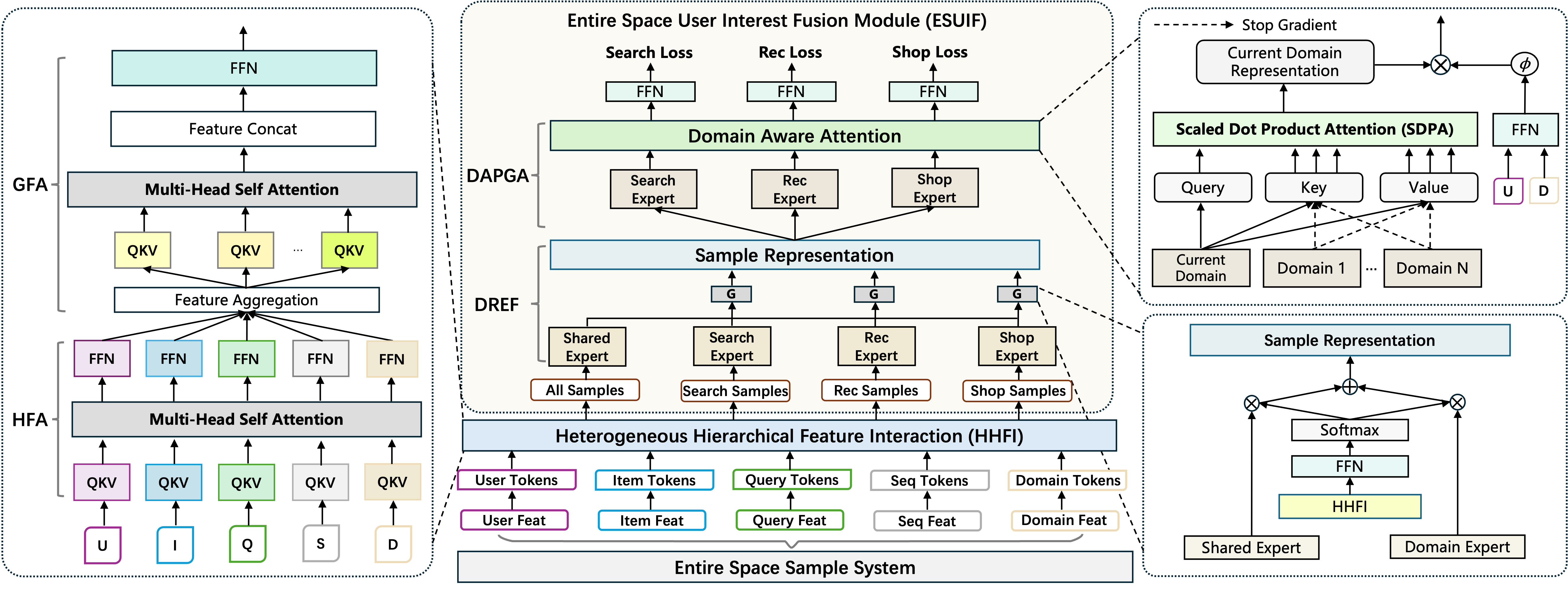}
\vspace{-14pt}
  \caption{Overview of Heterogeneous Hierarchical Sample Fusion Transformer (HHSFT) architecture. The HHSFT is designed to exploit model and data scaling in ranking systems through two primary stages. First, the Heterogeneous Hierarchical Feature Interaction (HHFI) stage employs heterogeneous feature attention layers with token-specific projection matrices and FFNs to preserve domain-unique distributions, followed by a global feature attention layer to capture high-order cross-domain interactions. Second, the Entire Space User Interest Fusion (ESUIF) stage comprises: Domain-Routed Expert Fusion (DREF) utilizes sample routing constraints to disentangle shared knowledge from domain-specific patterns; and Domain-Aware Personalized Gated Attention (DAPGA) adaptively modulates cross-domain information transfer via personalized gating.}
\vspace{-12pt}
\label{fig:hhsft}
\end{figure*}

\subsection{Heterogeneous Hierarchical Sample Fusion Transformer (HHSFT)}  

To fully exploit model and data scaling in ranking, we propose HHSFT, which captures high-order feature interactions through a hierarchical Transformer. The framework has two stages: \textit{Heterogeneous Hierarchical Feature Interaction} and \textit{Entire Space User Interest Fusion} (overview in \prettyref{fig:hhsft}).

\subsubsection{Heterogeneous Hierarchical Feature Interaction (HHFI)}\label{sec:hhfi}

We first partition input features into semantically coherent blocks (user attributes, item attributes, queries, user behavior) and map each block through an MLP to a unified token dimension, allowing heterogeneous features to be processed by the Transformer. However, standard Transformer encoders share projection matrices $\mathbf{W}^Q,\mathbf{W}^K,\mathbf{W}^V$ across all tokens. In e-commerce search, heterogeneous features (categorical user IDs, continuous prices, sequential behaviors) originate from significantly different semantic spaces, so shared projections lead to semantic confusion. We therefore propose a heterogeneous hierarchical feature interaction module with two parts:

\begin{itemize}[label={},leftmargin=0pt, noitemsep]
\item \textbf{Heterogeneous Feature Attention (HFA):} HFA equips each token with \textit{token-specific} QKV projections and FFN to preserve field-level semantics. In layer $l$, the $i$-th token representation $h_{i}^{(l)} \in \mathbb{R}^{d_H}$ is projected by dedicated weights $\mathbf{W}^{Q}_{i},\mathbf{W}^{K}_{i},\mathbf{W}^{V}_{i} \in \mathbb{R}^{d_H \times d_H}$:
\begin{align}
q_{i}^{(l)} = h_{i}^{(l)} \mathbf{W}^{Q}_{i}, \quad
k_{i}^{(l)} = h_{i}^{(l)} \mathbf{W}^{K}_{i}, \quad
v_{i}^{(l)} = h_{i}^{(l)} \mathbf{W}^{V}_{i}.
\end{align}
Standard multi-head self-attention then yields an updated token state $a_{i}^{(l)}$, which is passed through a block-specific FFN: $h_{i}^{(l+1)}=\text{FFN}_i(a_{i}^{(l)})$. Stacking $L_H$ such layers gives the final output $h_{i}^{L_H}$, which captures the unique semantic characteristics of each feature field while modeling cross-feature interactions.
\end{itemize}

\begin{itemize}[label={},leftmargin=0pt, noitemsep]
\item \textbf{Global Feature Attention (GFA):} Building upon the heterogeneous  attention layer, we utilize a composite projection mechanism to model high-level interactions between all tokens. The global feature attention layer obtains $m$ sets of  $\tilde{q}_j,\tilde{k}_j,\tilde{v}_j$ vectors with dimension $d_G$ through composite projections:
\begin{equation}
\tilde{k}_j = \text{concat}([h_{1}^{L_H}, \dots, h_{n}^{L_H}])  \tilde{\mathbf{W}}^K_j, 
\end{equation}
where $\tilde{\mathbf{W}}^K_j \in \mathbb{R}^{ (n \cdot d_H) \times  d_G}$ is the composite projection weight for the $j$-th token in the global feature attention layer.
The same projection strategy is applied to $\tilde{q}_j$ and $\tilde{v}_j$, ensuring symmetry in modeling interactions. Through multi-head self-attention fusion with composite projections, this layer learns comprehensive global interactions. Subsequently, all tokens are concatenated and passed through a Global FFN to produce an updated representation. By stacking the global feature attention modules up to $L_G$ layers, we obtain the final representation $\mathbf{Z}\in\mathbb{R}^{d_Z}$, which is utilized for downstream tasks.
\end{itemize}

\subsubsection{Entire Space User Interest Fusion (ESUIF)}
Naively mixing cross-domain samples (Section~\ref{subsec:feisou}) often induces severe negative transfer. ESUIF instead learns unified user-interest representations through two complementary components.

\begin{itemize}[label={},leftmargin=0pt, noitemsep]
\item \textbf{Domain-Routed Expert Fusion (DREF):}
\label{sec:moe_layer}
Under the multi-domain setting, user interests contain both transferable cross-domain commonalities and domain-specific intents that must be disentangled. Unlike standard MoE~\cite{ma2018modeling,tang2020progressive} where all experts are updated by all samples, we introduce \textbf{sample routing constraints}: a domain-shared expert $f_{s}$ forwards every sample to learn stable cross-domain patterns, while each domain-specific expert $f_{d}$ forwards and receives gradients only from samples in its domain $X_{d}$, isolating cross-domain noise at the optimization level. A gating network adaptively fuses the two:
\begin{equation}
[\alpha_{s},\alpha_{d}]=\mathrm{softmax}(g_{\mathrm{moe}}({z})),
\end{equation}
\begin{equation}
{e}=\alpha_{s}\cdot f_s(z)+\alpha_{d}\cdot f_{d}(z),
\end{equation}
where $g_{\mathrm{moe}}$ denotes sample-level gating network to adaptively fuse shared and specific domain branches.

\end{itemize}

\begin{itemize}[label={},leftmargin=0pt, noitemsep]
\item\textbf{Domain-Aware Personalized Gated Attention (DAPGA):}
Since Taobao Search is the only deployed domain, we need a target-centric transfer of auxiliary-domain knowledge. We propose \textbf{Domain-Aware Attention}, which is \textit{forward-visible across all domains} but \textit{backward-isolated within each domain} (Figure~\ref{fig:hhsft}): each sample ${e}$ is forward-propagated through all domain experts to yield multi-perspective representations $O$, then fused via SDPA where the target-domain output $o^{cur}$ serves as the query and all-domain outputs as keys/values, conditioning knowledge transfer on the user's current domain mindset. To prevent noise gradients from cross-domain samples, we apply loss masking on cross-domain logits and gradient stopping on cross-domain K/V projections:
\begin{equation}
O=[o^{cur},\text{sg}(o^1),...,\text{sg}(o^N)],
\end{equation}
\begin{align}
q^{cur}=o^{cur} \mathbf{W^Q} ,\quad
K=O \mathbf{W^K}, \quad
V=O \mathbf{W^V},
\end{align}
where $\text{sg}(\cdot)$ denotes gradient stopping. The domain representation $\tilde{o}$ is obtained by this mechanism.
\end{itemize}

User feedback also depends on current domain intent intensity and interaction costs. Inspired by gated attention~\cite{shazeer2020glu, qiu2025gated}, we add a \textbf{Domain-Aware Personalized Gating}: a lightweight sigmoid gate over user-side static features $u$ and domain features $d$ produces a scaling vector ${\gamma}$ that adds personalization and domain awareness to the fused representation:
\begin{equation}
\boldsymbol{\gamma} = \phi\left( {W}[Emb_u; Emb_d] +{b} \right)
\end{equation}
where ${\phi}$ denotes the sigmoid function, hence $ {\gamma}\in (0,1)$. Subsequently, we perform element-wise multiplication between ${\gamma}$ and the domain representation $\tilde{\mathbf{o}}$:
\begin{equation}
\hat{\mathbf{o}} = \boldsymbol{\gamma} \odot \tilde{\mathbf{o}}.
\end{equation}

\subsubsection{Loss Function}
Each domain has its own output $\hat{y}^{(d)}$. We minimize the per-domain binary cross-entropy loss:
\begin{equation*}
  \mathcal{L}(\theta) = -\sum_{d\in\mathcal{D}}\sum_{(\mathbf{x}, y) \in \mathcal{S}_d}  \left[ y \log( \hat{y}^{(d)})+ (1-y) \log(1 - \hat{y}^{(d)}) \right].
\end{equation*}
Although trained on all domains $\mathcal{D}$, the model is deployed only for $d=\text{search}$, leveraging auxiliary domains for transfer while keeping target-domain optimization.

\subsection{Training and Deployment Optimization}\label{sec:po}

Deploying UniScale at industrial scale requires a co-design of the underlying infrastructure to handle ultra-large parameter volumes and real-time latency constraints. Training-side optimizations include: \textit{(1) High-Concurrency Data Reader} that decouples I/O pre-fetching from GPU execution; \textit{(2) Feature Pre-hashing} performed at the earliest pipeline stage to cut memory overhead from raw string features; \textit{(3) RDMA-enabled cluster communication} replacing TCP for parameter synchronization. Inference-side optimizations include: \textit{(1) FP16 quantization} to maximize GPU throughput; \textit{(2) Tile-level operator push-down} that integrates lightweight ops (e.g., scaling, bias) into the embedding-lookup stage; \textit{(3) a fused masked QKV attention kernel} consolidating projection, transposition, and masking into a single op. Notably, ES\textsuperscript{3} itself adds no online compute cost—it is a training-time data construction system, and the trained HHSFT model serves with the same architecture regardless of training-data composition. The only ES\textsuperscript{3} overhead is offline storage for the expanded samples, which is small relative to online compute and yields a high ROI given the observed business gains.

\section{Experiments}

\subsection{Experiment Settings}

\subsubsection{Datasets}
To evaluate HHSFT, we conducted both offline and online experiments on Taobao's e-commerce dataset, comprising billions of user-item interactions. We apply ES\textsuperscript{3} (Entire-Space Sample System) to scale the training data from search-exposed samples only to a comprehensive entire-space dataset. As summarized in Table~\ref{tab:es3}, all values are scaled relative to the baseline (search-exposed samples only). ``Requests'' denotes unique search requests; ``Samples'' refers to item-level samples; ``Click Pos.'' counts samples enriched with all-domain click signals. Arrows ($\uparrow$) highlight the stage-specific contributions of ES\textsuperscript{3} components. To align our offline evaluation with the online deployment context, we constructed a held-out test set comprising only samples from the search domain. All reported metrics are computed on this search-specific dataset, providing a direct measure of the model's performance on the target task.

\begin{table}[htbp]
\vspace{-6pt}
\centering
\small 
\caption{Impact of ES\textsuperscript{3} on Training Data Scale}
\setlength\tabcolsep{4pt}
\vspace{-8pt}
\begin{tabular*}{\columnwidth}{@{\extracolsep{\fill}}lccc@{}} 
\toprule
\multicolumn{1}{c}{\textbf{Stage} } & \textbf{Requests} & \textbf{Samples} & \textbf{Click Pos.} \\
\midrule
Baseline (Search-exposed-only) & 1.0$\times$ & 1.0$\times$ & 1.0$\times$ \\
+ Unexposed Expansion & 1.0$\times$ & 3.0$\times$$\uparrow$ & 1.0$\times$ \\
+ Hierarchical Label Attribution & 1.0$\times$ & 3.0$\times$ & 2.0$\times$$\uparrow$ \\
+ Cross-domain Searchification & 2.0$\times$$\uparrow$ & 5.0$\times$$\uparrow$ & 4.0$\times$$\uparrow$ \\
\bottomrule
\end{tabular*}
\label{tab:es3}
\vspace{-12pt}
\end{table}

\subsubsection{Baselines} We compare against the following widely recognized SOTA baselines: 
\textbf{DLRM-MLP}: which is the vanilla MLP for
feature crossing, serves as as the experiment baseline; \textbf{DCNv2}~\cite{wang2021dcn} is the representative feature interaction model.
\textbf{AutoInt}~\cite{song2019autoint}, \textbf{Hiformer}~\cite{gui2023hiformer} investigate Transformer-like architecture for ranking model.
\textbf{Wukong}~\cite{zhang2024wukong} and \textbf{Rankmixer}~\cite{zhu2025rankmixer} investigate the scaling law for the recommendation model.
\subsubsection{Evaluation Metrics}
\label{sec:Evaluation Metrics}
To evaluate model performance, we use \textbf{AUC} (Area Under the Curve), \textbf{GAUC} (Group-AUC, user-query-wise) and \textbf{Hitrate@5} (HR@5) which measures top-5 recommendation accuracy as the primary performance metrics. Following industry practice, an increase of 0.05\% can be regarded as a confident improvement.  We also report \textbf{model size} (Params, in millions) and \textbf{computational complexity} (TFLOPs) to evaluate efficiency-scalability tradeoffs. For online experiments, we use GMV as the business-centric metric to reflect practical impact.

\subsection{Overall Performance}

The main results are summarized in Table~\ref{tab:cmp_sota}. To isolate the architectural contribution from the data contribution, all baselines and the HHSFT row are trained on the identical search-exposed dataset; only the final \textit{HHSFT+ES\textsuperscript{3}} row adds the entire-space training data. We observe three key findings: \textbf{1) Superiority of Transformer-based Architectures.} Transformer-based models (AutoInt, HiFormer, HHSFT) consistently outperform traditional DNN and FM-based baselines (DCNv2, Wukong), confirming the efficacy of explicit attention for high-order feature interactions. \textbf{2) Efficacy of hierarchical feature interaction.} On the same base data, HHSFT achieves +0.82\% AUC and +0.62\% GAUC, substantially exceeding HiFormer (+0.54\%/+0.49\%) and Wukong (+0.21\%/+0.13\%) under a comparable parameter budget. This isolates the architectural contribution independent of any data advantage. \textbf{3) Synergy between architecture and sample scaling.} \textit{HHSFT+ES\textsuperscript{3}} adds another +0.32\% AUC over HHSFT, and this benefit is architecture-dependent: as shown in Section~\ref{sec:mdr}, the same sample expansion \textit{degrades} a standard backbone by $-0.43\%$ AUC. HHSFT's routing mechanism is therefore essential for converting heterogeneous data from interference into a usable information gain.

\begin{table}[h]
\vspace{-6pt}
\centering
\setlength\tabcolsep{4pt}
\small 
\caption{Performance comparison of ranking models (best values in bold)
}
\vspace{-8pt}
\scalebox{1}{
\begin{tabular}{c|cccc}
\hline
Model &  AUC &GAUC & Params(M) & TFLOPs \\
\hline
 DLRM-MLP (base) & - & - &15 &0.42 \\ 
 DCNv2 &  +0.08\% & +0.01\% &24 &0.65\\ 
AutoInt  & +0.26\% & +0.14\% &150 & 1.19\\
HiFormer  & +0.54\% & +0.49\% & 170& 1.98\\ 
Wukong  & +0.21\% & +0.13\% & 32& 0.94\\ 
Rankmixer  & +0.38\% & +0.32\% & 140& 1.93\\ 
\hline
HHSFT & +0.82\%&  +0.62\% & 300& 1.22\\ 
HHSFT+ ES³ & \textbf{+1.14\%}& \textbf{+0.86\%} & 300& 1.22\\ 
\hline
\end{tabular}}
\label{tab:cmp_sota}
\vspace{-12pt}
\end{table}

\subsection{Ablation Study}

\subsubsection{Heterogeneous Hierarchical Feature Interaction (HHFI) Module}\label{sec:as-hhft}

Table \ref{tab:cmp_arch} reports per-component AUC improvements over the DNN-MLP baseline. \textbf{1) Transformer backbone}: replacing MLP with a basic Transformer encoder already yields significant gains, confirming that self-attention outperforms MLP in feature-interaction modeling. \textbf{2) Heterogeneous attention}: token-specific QKV projections and FFNs further boost performance, validating that disentangling heterogeneous feature semantics mitigates representation ambiguity. \textbf{3) Hierarchical feature interaction}: a global feature attention layer captures high-order interactions beyond pairwise dependencies. Combined, HHFI delivers a total $+0.80\%$ AUC gain, confirming that each design choice contributes meaningfully.

\begin{table}[h]

\vspace{-2pt}
\centering
\small 
\setlength\tabcolsep{4pt}
\caption{Ablation on components of HHFI.
}
\vspace{-8pt}
\scalebox{1}{
\begin{tabular}{c|c|c|c|c}
\hline
Setting &  MLP & Transformer  & Heterogeneous Attention  & HHFI\\
\hline
  AUC gain  & - & +0.17\% & +0.41\% & +0.80\%\\
\hline
\end{tabular}}
\label{tab:cmp_arch}
\vspace{-12pt}
\end{table}

\subsubsection{Entire Space User Interest Fusion (ESUIF) Module}\label{sec:mdr}
To quantify the incremental contribution of each key component in ESUIF Module under entire space sample setting, we conduct detailed ablation experiments. Table \ref{tab:sft} presents the gains relative to the metrics introduced in section \ref{sec:Evaluation Metrics}, with key insights as follows: \textbf{1) HHFI backbone with Entire-Space data:} naively mixing cross-domain data under HHFI backbone yields significant metric decrease, confirming severe negative transfer phenomena. \textbf{2) Domain-Routed Expert Fusion (DREF):} Explicitly disentangling shared and domain-specific representations via hard routing effectively suppresses cross-domain noise, yielding gains over intra-domain training and validating structural separation of shared and domain-specific patterns. \textbf{3) Domain-Aware Personalized Gated Attention (DAPGA):} Adaptively modulates target-domain representations via personalized gating and attention mechanism. It aligns cross-domain signals to contextual intent delivering further metric improvements beyond DREF. The combined effect of ESUIF results in a total +0.32\%/+0.26\%/+0.44\% gains in each metric over the HHFI training solely with intra-domain data, demonstrating that each design choice meaningfully contributes to the overall performance gain.

\begin{table}[ht]

\vspace{-6pt}
\centering
\small 
\setlength\tabcolsep{4pt}
\caption{Ablation on components of ESUIF.} 
\vspace{-8pt}
\scalebox{1}{
\begin{tabular}{c|c|ccc}
\hline
Sample & Setting &  AUC & GAUC & HR@5\\
\hline
\multirow{2}{*}{search} & HHFI & -  & -  & -\\ 
 & HHFI+DREF+DAPGA & +0.02\%  & +0.01\%  & +0.02\%\\ 
\hline
\multirow{4}{*}{ES³} &HHFI & -0.43\% & -0.24\% & -4.76\%\\ 
 & HHFI+DREF & +0.19\% & +0.25\% & +0.39\% \\ 
 & HHFI+DAPGA & +0.22\% & +0.18\% & +0.17\% \\ 
 & HHFI+DREF+DAPGA & \textbf{+0.32\%} & \textbf{+0.26\%} & \textbf{+0.44\%} \\ 
\hline
\end{tabular}}
\label{tab:sft}
\vspace{-12pt}
\end{table}

\subsection{Scaling Behavior Validation}

\begin{figure}[!ht]
\vspace{-6pt}
    \centering
    \includegraphics[width=0.25\textwidth]{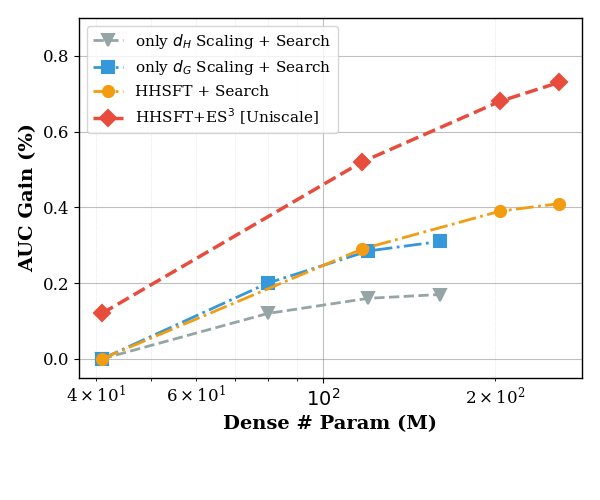}
\vspace{-20pt}
    \caption{AUC gain vs Dense Parameters Scale Ratio}
    \label{fig:scaling_exp} 
\vspace{-15pt}
\end{figure}

A central thesis of UniScale is that \textit{architectural scaling} and \textit{data scaling} must co-evolve. We validate this through systematic experiments. The base configuration uses dense architectural parameters (embeddings fixed): \textit{Heterogeneous feature attention} (1 layer, $d_H=1648$) and \textit{Global feature attention} (1 layer, $d_G=256$, $m=8$ tokens). We then scale model size and data volume independently and measure AUC gain in Figure~\ref{fig:scaling_exp}. We deliberately describe these observations as a controlled \textit{scaling trend} rather than a formal scaling law, since a 7$\times$ controlled sweep is insufficient to fit a power-law form; nevertheless the overall industrial expansion (40M$\rightarrow$300M, $\sim$7$\times$) is substantial. Two conclusions emerge:
\begin{itemize}[label={},leftmargin=0pt, noitemsep]
    \item \textbf{Model Scaling:} Under fixed search-domain data, scaling the global feature attention ($d_G$) yields substantially higher returns than scaling the heterogeneous feature attention ($d_H$) (+0.30\% vs.\ +0.17\% $\Delta$AUC at 300M parameters). This validates that HHSFT's hierarchical design strategically allocates capacity to high-order fusion components.

    \item \textbf{Synergistic Scaling Effect:} The gap between HHSFT trained on search-only data (HHSFT+Search) and HHSFT leveraging full heterogeneous data (HHSFT+ES\textsuperscript{3}) widens progressively with model scaling—from +0.12\% $\Delta$AUC at 40M to +0.32\% at 300M. We emphasize that both curves remain monotonically positive throughout: HHSFT+Search exhibits a flattening (positive but smaller) slope at 100M to 300M, while HHSFT+ES\textsuperscript{3} sustains a positive slope, so the diminishing-return point of model-only scaling is pushed significantly further by data co-scaling.
\end{itemize}

\noindent\textbf{Adoption Priority.} Based on the ablations, we recommend the following deployment order for practitioners with limited engineering bandwidth: \textit{(1) HHFI}---the core architectural change ($+0.80\%$ AUC over MLP, Table~\ref{tab:cmp_arch}), effective even without ES\textsuperscript{3}; \textit{(2) Unexposed expansion + Hierarchical Label Attribution}---relatively easy to land and immediately delivers $2\times$ positive labels and $3\times$ sample volume (Table~\ref{tab:es3}); \textit{(3) DREF + DAPGA + Searchification}---adds another $+0.32\%$ AUC over HHFI on entire-space data (Table~\ref{tab:sft}) and is most beneficial when extra-domain user behavior is substantially richer than in-domain behavior. A minimal viable system is HHFI together with the intra-domain expansion path; the remaining components compound the gain when cross-domain signal is rich.

\subsection{Online Deployment and Performance}
To validate industrial robustness, we conducted a 10-day A/B test on Taobao's search platform, allocating 5\% of production traffic to UniScale (HHSFT+ES\textsuperscript{3}) against the production DNN baseline. UniScale achieves a 1.70\% increase in purchase and a 2.04\% increase in GMV—statistically significant gains given Taobao's traffic scale. Notably, conversion metrics (harder to game than clicks) also improve, providing evidence that the cross-domain positives recovered by ES\textsuperscript{3} reflect genuine interest rather than mislabeled noise. The engineering optimizations in Section~\ref{sec:po} further reduce GPU inference cost by $\sim$55\% and training overhead by $\sim$40\%.

\section{Conclusion and Future Work}
We propose UniScale, a co-design framework that jointly optimizes data and architecture for industrial ranking model scaling, deployed on Taobao Search. ES\textsuperscript{3} scales data volume and mitigates selection bias, exposure bias, and cross-domain blind spots; HHSFT then absorbs the heterogeneous data through hierarchical feature interaction and entire-space user-interest fusion. Extensive offline and online experiments validate the synergistic co-design of data and architecture. In future work, we will pursue Trinity Scaling across sample, feature, and architecture dimensions.

\bibliographystyle{ACM-Reference-Format}
\bibliography{sample-base}

\end{document}